# PREDICTING ACADEMIC MAJOR OF STUDENTS USING BAYESIAN NETWORKS TO THE CASE OF IRAN

Shiva Asadianfam [1], Mahboubeh Shamsi [2], Sima Asadianfam [3]

[1] Department of Computer Engineering, Qom Branch, Islamic Azad University, Qom, Iran

[2] Department of Electrical & Computer Engineering, Qom University of Technology, Qom, Iran

[3] Department of Computer Engineering, Zanjan Branch, Islamic Azad University, Zanjan, Iran

## ABSTRACT

*In this study, which took place current year in the city of Maragheh in IRAN. Number of high school students in the fields of study: mathematics, Experimental Sciences, humanities, vocational, business and science were studied and compared. The purpose of this research is to predict the academic major of high school students using Bayesian networks. The effective factors have been used in academic major selection for the first time as an effective indicator of Bayesian networks. Evaluation of Impacts of indicators on each other, discretization data and processing them was performed by GeNIe. The proper course would be advised for students to continue their education.*

## 1. INTRODUCTION

Education is one of the major institutions of society are considered. Institutional arrangements can leave dramatic changes in many areas of social. Analyze the situation of progressive countries, we find that, they are paying attention to how education in recent decades has led to progress in various fields [1]. Dynamic and streamlined educational institution, an institution in which human resources are educate properly and according to the goals. The institutional knows needs, abilities and talents of people and its prosperity put forward regular programs consistent with academic standards [2]. Different nations due to of their efficient use of manpower and resources available study and careful planning. Because scientific research and careful planning that countries can use them to make the most of their potential in the global competition. Infrastructure planning for the institution of education is done [2].

In IRAN, to achieve the goals a high school education: "is to prepare students for employment or further education", there are the short-term and long-term planning, such as budgeting field of study in education. The new system of secondary education, there are five branch or field of study for further education. These fields of study are Technical & Vocational, Work & Knowledge, Mathematics & Physics, Experimental Sciences & Human Sciences.

Variety of fields in secondary education and higher education on the one hand, and a wide range of jobs in today's society and the relationship between education and job on the other will require students to their future educational and career based on personal interests and talents are correct planning. Today, education and employment are considered the most important questions in education and other social and economic aspects.

This paper is structured as follows. In Section 2, the background information required for the better understanding of the method presented in this paper is discussed. In Section 3, the empirical studies on Bayesian Networks are presented. In Section 4, material and methods are explained.Then, implementation of Bayesian Network by genie    is presented, and in Section 5, the results of the implementation are presented. Finally, in Section 6, the general conclusion is discussed.

## 2. BACKGROUND

Investigation on education issues and adaptation of results with predefined objectives can show the failure of the process of education and action to solve .What is evident in the current situation, the importance of fields students are in high school. High school graduates its first year based on interest, aptitude and performance in education and training in the assessment and preparation process changes are guided to one of the following paths: [3]

- Branch Of Theoretical (Mathematics And Physics - Experimental Sciences- Human Sciences)
- Technical And Vocational Categories
- Branch Of Work And Knowledge

But what on a student's choice of academic major influences, will also be important, select a field in the near future will bring jobs. In this section, will be discussed the factors influencing choice of academic major or students' choice of occupation.

### 2.1. The effective factors in choosing an academic major

Effective factors in the choice of academic major can be divided into two parts:
Individual factors: [4]

- Talent and mental ability
- Willingness and  interest
- Personality traits
- Knowledge and skills

- Academic records
- Personal needs
- Sex
- Personal experiences

Environmental factors: [4]

- Cultural Foundations
- Social position
- Geographic Location
- Economic situation
- industry changes
- Religion

- Family and Friends
- School
- Family
- Academic guidance

These are all hand in hand to the person to choose their field of study, but given the subject matter here is trying that important factors to be considered and given these factors, academic major must be selected for further study.

## 2.2. Bayesian Networks

If ($a_1$, $a_2$, ... , $a_n$) be a set of random variables, Bayesian Networks can express the probability of any combination of them. Based on Bayesian Networks, Bayesian rule that can be expressed as (1).

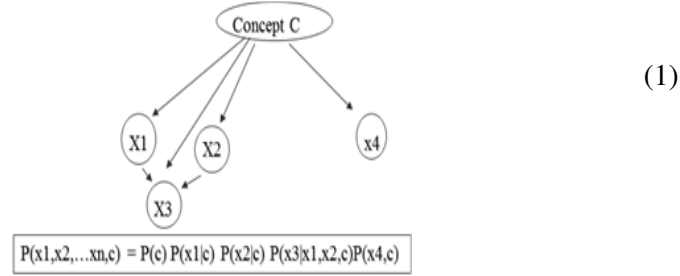

(1)

This network is a natural tool to address two problems in engineering and applied mathematics that provide uncertainty and complexity of the issues.

A Bayesian Network is called Belief Network or directional probabilities networks, a graph possibilities show the random variables and their dependencies. The network nodes represent random variables and arcs represent dependencies between the random variables with the conditional probability. The network is a graph with directional and no loop. so, All edges are directional and there are no loops [5, 6].

## 3. RELATED WORK

There is a wide range of application for Bayesian Network. In the paper [7], a Bayesian Network model was built to represent the probability distribution of each factor and how they affect defects, considering strong or weak correlations are existed between individual metric attributes. The software metrics and build a Bayesian Network model for defect prediction.

Khlifia Jayech and Mohamed Ali Mahjoub propose to use new approach combining distance tangent, k-means algorithm and Bayesian network for image classification in [8]. First, they use the technique of tangent distance to calculate several tangent spaces representing the same image. The objective is to reduce the error in the classification phase. Second, they cut the image in a whole of blocks. For each block, they compute a vector of descriptors. So, they use K-means to cluster the low-level features including color and texture information to build a vector of labels for each image. Finally, they apply five variants of Bayesian networks classifiers to classify the image of faces using the vector of labels.

In the paper [9] proposed a new framework for discovering interactions between genes based on multiple expression measurements.This framework builds on the use of Bayesian Networks for representing statistical dependencies.

In the paper [10], Pablo Felgaer and Paola Britos define an automatic learning method that optimizes the Bayesian Networks applied to classification, using a hybrid method of learning that combines the advantages of the induction techniques of the decision trees (TDIDT-C4.5) with those of the Bayesian Networks. The resulting method is applied to prediction in health domain.

In [11], the aim of study is model development for financial distress prediction of listed companies in Tehran stocks exchange (TSE) using Bayesian networks (BNs). it can be an evidence that the financial statements of companies have information content. With respect to the remainder variables in developed models in this research we find firms that have lower profitability and have more long term liabilities and have lower liquidity are more in risk of financial distress.

## 4. MATERIAL AND METHODS

### 4.1. Random Variables And Data Collection

The study was conducted at Maragheh city In IRAN this year, number of students of secondary school in the academic major, Technical And Vocational, Work And Knowledge, Mathematics And Physics, Experimental Sciences And Human Sciences via questionnaire were compared to the parameters and their values are shown in Table1.

Table 1 -Random variables used in the Bayesian Network

| Random variable name | Description | Values |
|---|---|---|
| High_school_score | junior high school grades | |
| Middle_school_score | School grades | |
| Not_come_score | Hush threshold score | |
| University | Way into university | |
| Paremt_major | Parental field of study | |
| Parent_guide | Parents Guidelines | Very much |
| Teachers_guide | Teachers Guidelines | |
| Manager_guide | Principal Guidelines | Much |
| Adviser_guide | Consultant Guidelines | |
| Firend_advise | Friend Recommend | Low |
| Weeky_plan | Program Overview of Introduction jobs | |
| Kinfolk_major | field of study relatives | very low |
| Job | Jobs of the Future | |
| Sociaty_requirment | Awareness of community needs | |
| Sociaty_lookout | view than field of study | |
| Social_position | The job social status | |
| Salary | The job salary and wage | |
| Tendency | Interest | |

### 4.2. Form of Bayesian Networks

In this paper, the software GeNIe is used to creat a Bayesian Network. First, identify and show relationships between random variables in the network. Given the important factors in choice of academic major with the data obtained from the research that has been done in the city.Also, consult with an academic advisor are obtained relationships between variables. The frequency and the percentage effect of each agent is calculated according to Quad ratings in students' choice academic major for each question and the average effect of each factor according to the number of students choices each of the four levels range from very low ,low, high, and very high have been calculated.   Finally, after the creation of networks and test them, model is presented, which consists of 19 nodes and 18 links is shown in Figure 1.

Due to the weight given to each of Quad spectra and using the following formula, the frequency of each of the variables measured and the mean impact of factor was obtained that the full impact factors directly connected with a direct link to the target [9].

frequency percentage every spectrum of factors (very much, much, low, very low) $= \dfrac{f_{(i)}}{100}$

$$\text{average effect of each factor} = \frac{f_{i(1)} \times 1 + f_{i(2)} \times 2 + f_{i(3)} \times 3 + f_{i(4)} \times 4}{100}$$

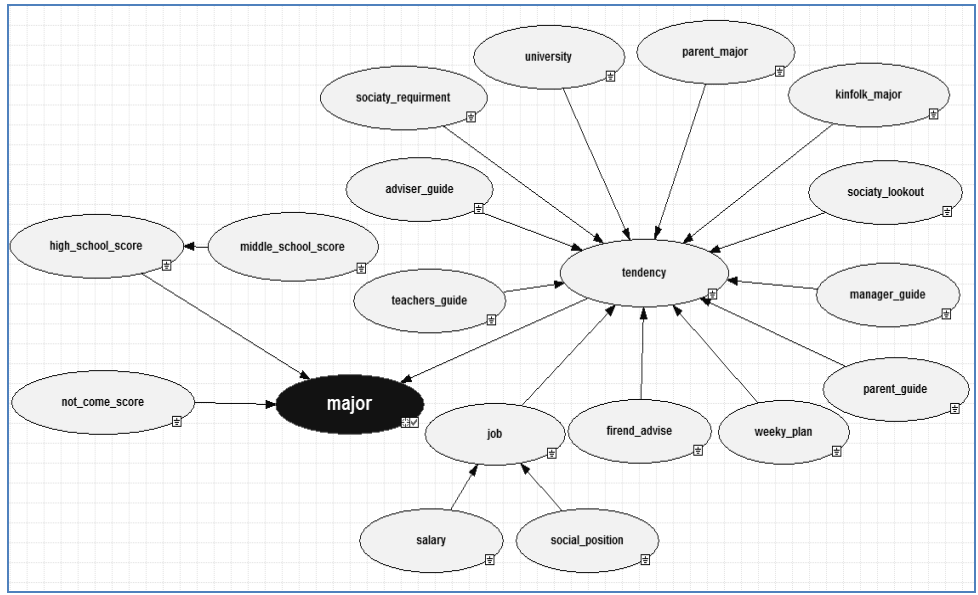

Figure1- Bayesian network used to predict the field of study

## 4.3. Chart Of Random Variables

According to the data collected and discretization of data, diagrams of random variables were drawn that two samples are shown in Figure 2 and 3.

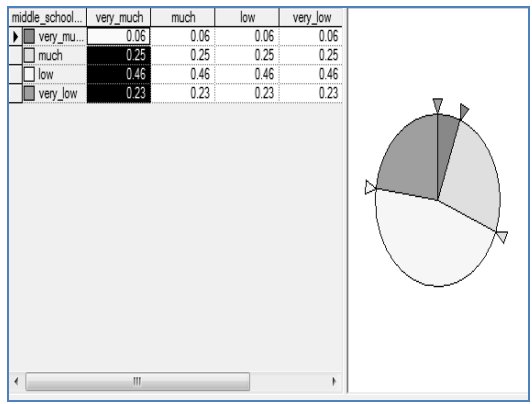

Figure3-Diagram of data frequency of parent_major

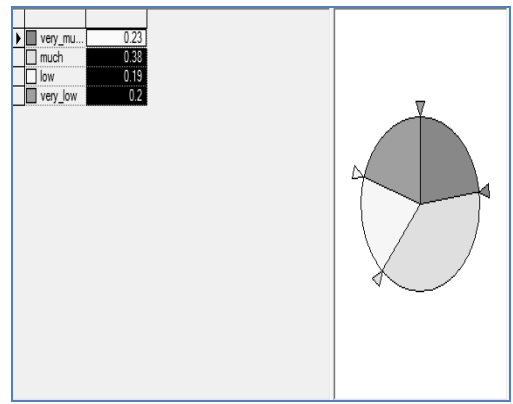

Figure2- Diagram of data frequency of jjob

## 4.4. Implementation of Bayesian network by GeNIe

Bayesian Network was drawn after identifying relationships between variables and calculate the probability of each node in the network and the formation of Conditional Probability Tables (CPT). Finally, for network inference was used algorithms as different as Logic Sampling, Likelihood sampling and EPIS Sampling. The Bayesian Network are shown in Figure 4.

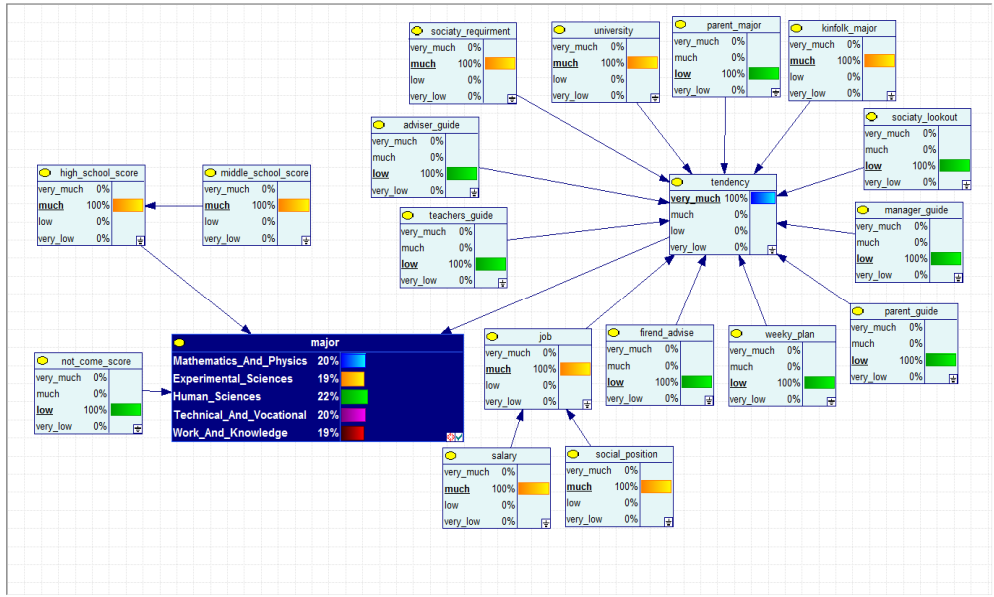

Figure 4- Bayesian network of secondary school students using GeNIe

# 5. EXPERIMENTAL RESULT ANALYSIS

Using Bayesian Networks obtained, the impact of various factors in field of study can be examined and offered the appropriate academic major. Finaly, the processing was carried out using the above-mentioned algorithms, the results are shown in Table 2. Thus we can say that for test data, the network has the ability to correct diagnosis about 70%. In Bayesian Networks, as our input data increase, the output of the network show us the results are more likely.

Table 2 -Results of the Bayesian Network using three different algorithms

| Algorithm | 1 | 2 | 3 | 4 | 5 | 6 | 7 | 8 | 9 |
|---|---|---|---|---|---|---|---|---|---|
| Likelihood Sampling | %70 | %70 | %70 | %70 | %70 | %72 | %70 | %70 | %78 |
| Logic Sampling | %77 | %78 | %77 | %77 | %79 | %78 | %79 | %78 | %78 |
| EPIS Sampling | %70 | %73 | %75 | %70 | %73 | %76 | %70 | %70 | %73 |

# 6. CONCLUSIONS

Bayesian Networks are widely used in detection and prognosis of diseases. The purpose of this study is to predict the academic major selection for high school students using Bayesian networks. The effective factors have been used in academic major selection for the first time as an effective indicator of Bayesian networks According to the survey results of Bayesian Network outputs with different algorithms using various data, we find that the better outcome for Logic Sampling algorithm of data.